# Geometry-assisted topological transitions in spin interferometry

[1]M. Wang, [2]H. Saarikoski, [3]A. A. Reynoso, [4]J. P. Baltanás, [4,5]D. Frustaglia[*], and [1,6,7]J. Nitta[**]

[1]*Department of Materials Science, Tohoku University, Sendai, 980-8579, Japan.*
[2]*RIKEN Center for Emergent Matter Science (CEMS), Wako Saitama 351-0198, Japan.*
[3]*INN-CONICET, Centro Atómico Bariloche, 8400 Bariloche, Argentina.*
[4]*Departamento de Física Aplicada II, Universidad de Sevilla, E-41012, Sevilla, Spain.*
[5]*Freiburg Institute for Advanced Studies (FRIAS), Albert-Ludwigs Universität Freiburg, D-79104 Freiburg, Germany.*
[6]*Center for Spintronics Research Network, Tohoku University, Sendai, 980-8577, Japan.*
[7]*Organization for Advanced Studies, Center for Science and Innovation in Spintronics (Core Research Cluster), Tohoku University, Sendai, 980-8577, Japan.*

[*]*frustaglia@us.es*
[**]*nitta@material.tohoku.ac.jp*

We identify a series of topological transitions occurring in electronic spin transport when manipulating spin-guiding fields controlled by the geometric shape of mesoscopic interferometers. They manifest as distinct inversions of the interference pattern in quantum conductance experiments. We establish that Rashba square loops develop weak-(anti)localization transitions (absent in other geometries as Rashba ring loops) as an in-plane Zeeman field is applied. These transitions, boosted by non-adiabatic spin scattering, prove to have a topological interpretation in terms of winding numbers characterizing the structure of spin modes in the Bloch sphere.

Spin-orbit (SO) interaction emerges as a vector potential leading to momentum-dependent magnetic textures that govern the evolution of itinerant spins in mesoscopic systems [1]. The resulting spin dynamics can exhibit a series of geometric and topological properties of significant interest as the Aharonov-Casher (AC) effect [2], the electromagnetic dual of the Aharonov-Bohm (AB) effect [3]. Ring-shaped Rashba SO interferometers have been shown to develop radial magnetic textures that can be manipulated by applying additional in-plane Zeeman fields [4]. At a critical point, the hybrid Rashba/Zeeman field undergoes a topological transition from a rotating texture (dominated by the radial Rashba component) to an oscillating one (dominated by the uniform Zeeman component) leaving an imprint on the spin dynamics with remarkable consequences in quantum transport experiments. This was first discussed by Lyanda-Geller [5] in the limit of adiabatic spin transport, where the fields are sufficiently strong to force the itinerant spins to stay locally aligned with the Rashba/Zeeman field texture during transport [6]. In this limit, field and spin textures share identical topological features captured by the itinerant spins in the form of a geometric Berry phase [7] that manifests as a spin-dependent magnetic flux in the quantum interference of spin carriers (recalling the AB effect [9]-[16]) that can be switched on/off by the in-plane Zeeman field [8]. We have shown that topological spin phase transitions are still possible in ring-shaped interferometers away from the adiabatic limit [17]: The transition is determined by the field's topology through an effective Berry phase related to the winding number of the spin states around the poles of the Bloch sphere. Yet, the experimental realization is challenging since the required in-plane Zeeman fields are strong, giving rise to spin-induced dephasing [18] that ruins spin interference as the transition point is approached [4].

Recently, the possibility of spin reorientation by geometric means has been considered after noticing that out-of-plane spin components develop in curved Rashba SO channels [19,20]. An early proposal for the geometric manipulation of electron spins was discussed in polygonal interferometers [21]-[22], where the vertices act as spin-scattering centers that hinder the emergence of AC and geometric phases. This suggests that sharp curvatures could soften the stiffness of Rashba-driven spin textures under the action of in-plane Zeeman fields, turning topological spin phase transitions possible at much weaker field strengths.

Here, we report transport simulations and experiments in square-shaped Rashba spin interferometers. Our simulations show that the topological properties of the spin states undergo a transition as an in-plane Zeeman field is applied. The topological transitions manifest as a sign reversal of spin interference leading to a distinctive weak-(anti)localization pattern for weak

Zeeman fields that contrasts with the case of ring-shaped interferometers. We demonstrate this response in a transport experiment with semiconductor square-shaped spin interferometers, establishing the existence of spin-texture topological transitions assisted by the polygonal geometry of the SO channels.

**One-dimensional (1D) transport simulations–** 1D Rashba SO wires along a generic direction $\hat{\gamma}$ in the *xy* plane are modeled by the Hamiltonian [21]

$$H_\gamma = \frac{p_\gamma^2}{2m^*} + \frac{\alpha}{\hbar} p_\gamma (\hat{\gamma} \times \hat{z}) \cdot \boldsymbol{\sigma}, \qquad (1)$$

with $p_\gamma$ the linear momentum of the spin carriers, $\alpha$ the Rashba SO strength, and $\boldsymbol{\sigma}$ the Pauli matrix vector. The second term in (1) appears as an effective in-plane magnetic field $\boldsymbol{B}_{SO} = \frac{2\alpha}{\hbar g \mu_B} p_\gamma (\hat{\gamma} \times \hat{z})$ coupled to the itinerant spin, with $\mu_B$ the Bohr magneton and $g$ the g-factor. Notice that $\boldsymbol{B}_{SO}$ is perpendicular to the wire's direction $\hat{\gamma}$ and inverts its sign for counter-propagating carriers due to the factor $p_\gamma$. Polygonal Rashba loops with *N* sides and perimeter *P* can be built by arranging segments of length $l = P/N$ modeled by (1), leading to effective-field discontinuities at the vertices with a significant effect on spin dynamics. Ring-shaped loops are described by taking the limit *N>>1* with constant *P*, where a radial effective field emerges and the discontinuities disappear [23] (see Fig. S2). Moreover, we introduce an in-plane uniform magnetic field $\boldsymbol{B}_Z$ along the *x*-axis that interacts with the itinerant spins through a Zeeman-coupling term $\frac{g\mu_B}{2} B_Z \sigma_x$ to be added in (1). The Rashba SO and Zeeman coupling strengths are quantified in terms of the corresponding spin precession angles during propagation along the loop's perimeter, $k_{SO}P$ and $k_Z P$, with $k_{SO} = \frac{\alpha m^*}{\hbar^2}$, $k_Z = \frac{g\mu_B B_Z m^*}{2\hbar^2 k_F}$, and $k_F$ the Fermi wave number. The hybrid Rashba/Zeeman field texture changes topology at the turning point $k_z P = k_{SO} P$.

We calculate the zero-temperature quantum conductance of 1D ring- and square-shaped loops with source/drain contact leads along the *x*-axis by applying the Landauer-Büttiker formalism, which identifies conductance with transmission. Moreover, we apply a semiclassical approach (valid when the system's size $P$ is much larger than the Fermi wavelength $\lambda_F$, in coincidence with our experimental conditions) where the quantum transmission results from the interference of classical paths contributing to the transport of carriers through the loop. Along their way, spin carriers collect different quantum phases according to their particular path, spin species, and fields experienced. In experiments where disorder and/or sample average is relevant, one finds that only pairs of paths with the same geometric length contribute significantly to quantum interference (contributions from other paths simply average out). These correspond to time-reversal (TR) orbital paths starting and ending at the source contact. In the absence of fields, TR paths produce constructive interference that reinforce backscattering and leads to magnetoconductance minima, something referred to as weak localization (WL). The introduction of SO fields can revert this effect by means of spin AC phases contributing to destructive TR-path interference, producing magnetoconductance maxima known as weak antilocalization (WAL). Large Zeeman fields can restore WL by "freezing" the spin dynamics [18]. Hence, WL/WAL signals in the presence SO and Zeeman fields reflect the response of the spin degree of freedom. See Supplementary Material for a technical summary.

Figures 1(a) and 1(b) show the conductance of 1D ring- and square-shaped loops calculated from TR-path interference at the lowest order, i.e., by considering only single-winding (counter)clockwise paths (see insets, dotted arrows), as a function of $k_{SO}P$ and $k_ZP$. The contrast found between the interference patterns of Figs. 1(a) and 1(b) indicates that spin phases $\phi_s$ contributing to WAL (blue zones) develop quite differently in rings and squares. Spin phases are dominated by the expectation value of the spin Hamiltonian $H_s = \frac{g\mu_B}{2}[\boldsymbol{B}_{SO}(\ell) + \boldsymbol{B}_Z] \cdot \boldsymbol{\sigma}$ over the itinerant-spin eigenstates $|\chi_s(\ell)\rangle$, $\phi_s \sim \int_0^P \langle \chi_s(\ell)|H_s|\chi_s(\ell)\rangle\, d\ell$, i.e., by the projection of the spin texture $\hat{\boldsymbol{s}}(\ell) = \langle \chi_s(\ell)|\boldsymbol{\sigma}|\chi_s(\ell)\rangle$ over the field texture $\boldsymbol{B}(\ell) = \boldsymbol{B}_{SO}(\ell) + \boldsymbol{B}_Z$, where $\ell$ parametrizes the loop's perimeter. Moreover, spin and field textures can be characterized topologically in terms of (integer) winding numbers around the $z$-axis defined as $\omega_{s/B} = \frac{1}{2\pi}\int_0^P d\ell \left(\hat{\boldsymbol{n}}_{s/B} \times \frac{d\hat{\boldsymbol{n}}_{s/B}}{d\ell}\right) \cdot \hat{\boldsymbol{z}}$, with $\hat{\boldsymbol{n}}_{s/B}(\ell)$ the normalized $xy$-projections of $\hat{\boldsymbol{s}}(\ell)$ and $\boldsymbol{B}(\ell)$, respectively, where $\omega_{s/B} = 0, \pm 1$ in the cases studied here [24]. WAL arises only when the spin phases are sufficiently large (otherwise, WL prevails). Figures 1 and 2 show that this happens when the field and spin textures share the same topology ($\omega_s = \omega_B$), i.e., when the local projection $\hat{\boldsymbol{s}}(\ell) \cdot \boldsymbol{B}(\ell)$ has a definite sign all over the loop's perimeter. We elaborate on this along the following paragraphs.

In Rashba rings, it has been shown that spin and field textures are well correlated even in the presence of Zeeman fields (away from the critical point $k_zP = k_{SO}P$) [4,17,25,26]. This is illustrated in Fig. 1(c), where we plot the spin winding number $|\omega_s|$. For field textures dominated by the radial SO component ($k_zP < k_{SO}P$, $|\omega_B| = 1$) the spin textures develop a finite $\omega_s$ (black zone) corresponding to cone-like trajectories of $\hat{\boldsymbol{s}}(\ell)$ winding around the $z$-axis in the Bloch sphere, see Fig. 2(A). As for the conductance, Fig. 1(a), it develops regular

AC interference bands with a period of $\pi$ in $k_{SO}P$ units. In the adiabatic limit $\frac{k_{SO}P}{\pi} \gg 1$, spin and field textures eventually coincide within the ring's plane. Moreover, the AC interference bands bend as the Zeeman field is introduced. This shift has been experimentally observed and explained as a spin geometric phase effect [4]. Figure 1(c) shows that changing the spin winding $\omega_s$ (white zone) requires large Zeeman fields beyond the critical point ($k_z P > k_{SO} P$) that changes the field winding $\omega_B$ as well, see Fig. 2(B).

In Rashba squares, Fig. 1(b), we find that the AC oscillation period doubles to $2\pi$ in $k_{SO}P$ units with respect to rings, in agreement with previous works for Rashba squares without Zeeman fields [21]. The discontinuity of the Rashba SO field at the corners turns adiabatic spin evolution impossible even for large $k_{SO}P$, hindering the development of AC phases. Spin eigenstates stay away from the loop's plane, forming complex spin textures sensitive to Zeeman-field perturbations, as illustrated by Figs. 2(C) and 2(D). The checkerboard-like pattern emerging in Fig. 1(b) as $k_z P$ increases shows interference fringes fully reversed by a weak Zeeman field on a scale $\pi$ in $k_z P$ units. This pattern– contrasting with the interference bands of Fig. 1(a) for rings– is due to topological changes in the spin texture. Indeed, Fig. 1(d) shows that the spin winding $\omega_s$ is fully correlated with the interference pattern of Fig. 1(b): Conductance maxima (WAL) in Fig. 1(b) (blue zones) are the consequence of large spin phases that can only be produced by spin textures with finite windings– and significant projection on the equally winding field texture, see Fig. 2(C)– appearing as dark zones in Fig. 1(d). A weak Zeeman field introduces major distortions on the highly non-adiabatic spin textures, producing a decorrelation between spin- and field-texture windings– see Fig. 2(D)– that impede the gathering of spin phases contributing to WAL, therefore restoring WL. This demonstrates that spin-texture topological transitions are accessible in square loops by using weak Zeeman fields that do not modify the topology of the field texture, in contrast to what is observed in rings. See Supplementary Material for an extended discussion.

**Experiment and discussion**– Our experimental setup consists of an array of Rashba square-shaped spin interferometers, which we use to investigate the topological transition induced by an in-plane Zeeman field. This configuration allows for a clear observation of the spin interference thanks to the ensemble averaging naturally provided by the sample.

We employed an InGaAs quantum well (QW) epitaxially grown on a InP (001) substrate. The detailed layer structure of the QW consists, from the bottom, of $In_{0.52}Al_{0.48}As$ (200 nm, buffer layer)/$In_{0.52}Al_{0.48}As$ (6 nm, carrier supply layer; Si-doping concentration of 4 x $10^{18}$ cm$^{-3}$)

/In$_{0.52}$Al$_{0.48}$As (15 nm, spacer layer) /In$_{0.53}$Ga$_{0.47}$As (2.5 nm, QW)/In$_{0.73}$Ga$_{0.27}$As (10 nm, QW) / In$_{0.53}$Ga$_{0.47}$As (2.5 nm, QW)/ InP (5 nm, stopper layer) /In$_{0.52}$Al$_{0.48}$As (20 nm, barrier layer)/AlAs (1.5 nm, barrier layer) /In$_{0.52}$Al$_{0.48}$As(5 nm, cap layer). The carrier density dependence of the Rashba SO parameter $\alpha$ is obtained from the analysis of the beating patterns appearing in the gate voltage dependence of Shubnikov-de Haas oscillations. The Rashba SO parameter $\alpha$ is electrically controlled from $\alpha$= -1.17 peVm at $N_s$ = 2.0 x 10$^{12}$ cm$^{-2}$ ($V_g$= -2 V) to $\alpha$= -3.62 peVm at $N_s$ = 1.26 x 10$^{12}$ cm$^{-2}$ ($V_g$= -5 V).

A top-gate-attached 40×40 square loop array with square side length $l$ = 700 nm was fabricated by electron beam lithography and reactive ion etching. A laser scanning microscope image is shown in Fig. 3(a). The square loop array was covered with a 200-nm thick Al$_2$O$_3$ insulator made by atomic layer deposition and a Cr/Au top gate electrode in order to tune the Rashba SO strength $\alpha$ [27], [28]. All the measurements were performed at a temperature of 1.7 K. The in-plane magnetic field $\boldsymbol{B}_Z$ points along the squares diagonal.

We focus our study on the $B_Z$ dependence of the Al'tshuler-Aronov-Spivak (AAS) [29] oscillation amplitude due to TR-path interference in the presence of a magnetic flux produced by a perpendicular field $B_\perp$. The AAS amplitude at $B_\perp = 0$ as a function of top-gate voltages reflects exclusively the phase contribution from the spin part of the wave function. The sign reversal of the AAS oscillations indicates that the constructive spin interference switches to destructive interference (WAL), or vice versa (WL restoration).

We measured the magnetoresistace (MR) by applying a weak $B_\perp$ (few mT strength) as a function of the gate voltage $V_g$ that controls $\alpha$. A series of MR measurements was performed for different values of the $B_Z$, running from $B_Z$= 0 T to $B_Z$= 2.5 T in steps of $\Delta B_Z$= 0.5 T. Figures 3(b)- (g) shows the contour color plots of the MR as a function of $V_g$ for different $B_Z$. The insets in Figs. 3(b)-(g) contain MR data at $V_g$= -3.6 V, showing AAS oscillations with period $\Delta B_\perp = (h/2e)l^2$ due to TR-path interference. The MR amplitude at $B_\perp = 0$ as a function of $V_g$ corresponds to the AC spin interference [30]-[31] (definite AC modulations by $V_g$ are better seen in Figs. S4 and S5). As discussed in Fig. 1(b), the AC oscillation period in the square loop is two times larger than the AC period of the ring-shaped spin interferometer. This is the reason why a whole period of the AC oscillation is not observed in the present sample (-1.17 peVm < $\alpha$ < -3.62 peVm).

By increasing $B_Z$, the interference amplitudes in Figs. 3(b)-(g) weaken gradually due to spin-induced dephasing [18] all over the gate voltage range. Still, a definite structure emerges as a function of $B_Z$. Around $B_Z$ = 1 T, the AAS amplitude vanishes at $V_g$= -3.6 V. Further increase in $B_Z$ (> 1.5 T) turns the AAS oscillations on again, but with a reversed AAS interference pattern. The sign reversal of the AAS oscillations corresponds to the transition from a destructive spin interference (WAL) to a constructive one (WL). This is clearly seen in Fig. 4(a). It is interesting to compare this response with the corresponding one observed in a ring array, shown in Fig. 4(b) for a fixed $V_g$= -4.6 V (corresponding to $\alpha$ = -2.9 peVm ). The sample (ring radius 600 nm) is the same one used in [4]. No inversion of the AAS oscillation pattern as a function of $B_Z$ is observed in this case, so that the topological transition from WAL to WL never occurs in the studied gate voltage region. These results are consistent with the calculations shown in Figs. 1(a) and 1(b) (striped vs. checkerboard patterns) based on the 1D model. The observed AAS oscillation reversal in the square loop array is attributed to the topological transition on the spin textures as discussed in Figs. 1 and 2.

In a previous work, we demonstrated [17] that the topological transition in a ring happens when the in-plane Zeeman field equals the Rashba SO field (namely, a topological transition in the spin texture requires a topological transition in field texture), which is given by $B_{SO}$ = $2\alpha k_F/g\mu_B$. The corresponding in-plane Zeeman field is estimated to be $B_Z$= 6.7 T with $\alpha$= -2.5 peVm, $N_s$ = 1.6 x $10^{12}$ cm$^{-2}$, and $g$ = 4. The experimentally observed transition field in the square loop array is around $B_Z$= 1.5 T, which is much weaker than the expected in the ring-shaped interferometer and does not require a topological change in the hybrid Rashba/Zeeman field texture. This is the main result of our work.

**Two-dimensional (2D) transport simulations–** The 1D model is useful to gain physical insight into the role played by the shape of the SO interference loop. However, it does not clarify how the topological transition is affected by scattering and multi-mode transport. Therefore, we resort to fully-quantum 2D numerical simulations of disordered multi-mode square loops accounting for more realistic conditions. We implement a tight-binding model that is solved by using a recursive Green's function method and the Kwant code [32] with parameters fitting those in the InGaAs QW used in our experiment (see Supplementary Material for technical details). The numerical result shown in Fig. S6(a) reproduces the experimental findings of Fig. 4(a) rather well. We further notice that the critical $B_Z$ producing the topological transition depends on the mean free path as shown in Fig. S6(b). This finding is compatible with previous results showing that disorder contributes to the development of non-adiabatic spin dynamics [6], boosting the effects of SO-field discontinuities at the square's corners.

**Conclusions–** The interference pattern found in our transport experiments with a Rashba square-loop array shows transitions from WAL to WL by introducing weak in-plane Zeeman fields. This experimental result matches the checkerboard-like pattern of the conductance as a function of the Rashba and Zeeman fields predicted by the theory, which has a topological interpretation in terms of spin-texture winding numbers (fully correlated with WAL-WL zones). The transitions found in the square-shaped interferometer differ significantly from the striped pattern found in the ring-shaped ones, demonstrating the distinct structure of spin textures and phases. In Rashba rings, topological transitions in spin textures require large Zeeman fields (indeed, a topological transition in the field texture is needed). The situation is very different for Rashba squares, where small Zeeman fields can produce the transition due to the role played by the squares corners as spin scattering centers. Additional sources of SO scattering as disorder appear to boost this effect.

Finally, we notice that the periodic (checkerboard-like) structure of the conductance and winding patterns for large Rashba strengths in squares, Figs. 1(b) and 1(d), suggests the existence of Brillouin-like zones for spin textures along the Rashba and Zeeman parameter spaces. This opens up the possibility of characterizing the reported topological transitions in terms of Chern numbers [33].

**Acknowledgements:**
This work was supported by the Japan Society for the Promotion of Science through Grant-in-Aid for Specially Promoted Research No. H1505699, Grant-in-Aid for Scientific Research (C) No. 17K05510, Grant-in-Aid for Innovative Areas No. JP15K21717, and Core-to-Core Program, and by Projects No. FIS2014-53385-P and No. FIS2017-86478-P (MINECO/FEDER, Spain). D.F. acknowledges additional support from the Marie Sklodowska-Curie Grant Agreement No. 754340 (EU/H2020). The 2D simulations were calculated using the HOKUSAI system provided by Advanced Center for Computing and Communication (ACCC) at RIKEN. J.N. and M.W. are grateful to Makoto Kohda for valuable discussions.

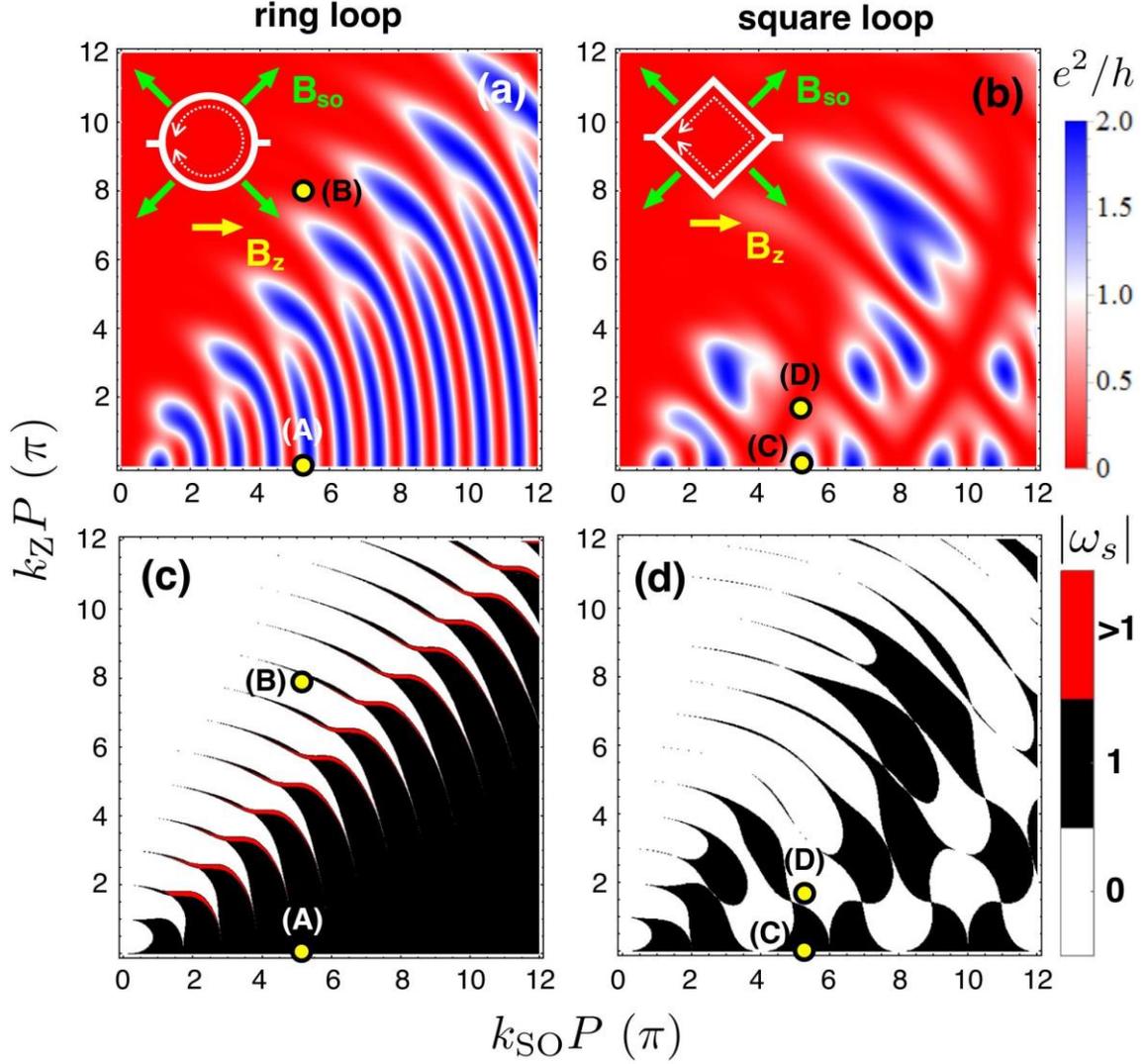

Fig. 1. Linear conductance of ring (a) and square (b) loops by TR-path interference (red: WL; blue: WAL) and winding number of spin textures, $|\omega_s|$, in ring (c) and square (b) loops as a function of the Rashba SO and in-plane Zeeman field strengths. Field and spin textures corresponding to points (A), (B), (C) and (D) are shown in Fig. 2. For dominating Rashba strengths (field winding $|\omega_B| = 1$), conductance maxima (blue) take place only when $|\omega_s| = 1$ (dark). Checkerboard-like patterns in (b) and (d) are indicative of spin-texture topological transitions for weak Zeeman fields in squares, in contrast to rings. Rashba SO field textures (insets) correspond to counterclockwise itinerant spins ($\boldsymbol{B}_{SO}$ flips over for clockwise spin carriers). Interference fringes in (a) and (b) disappear for dominating Zeeman fields due to a frozen spin dynamics. Red zones in (c) ($|\omega_s| > 1$) are due to resonances discussed in [17] and [25].

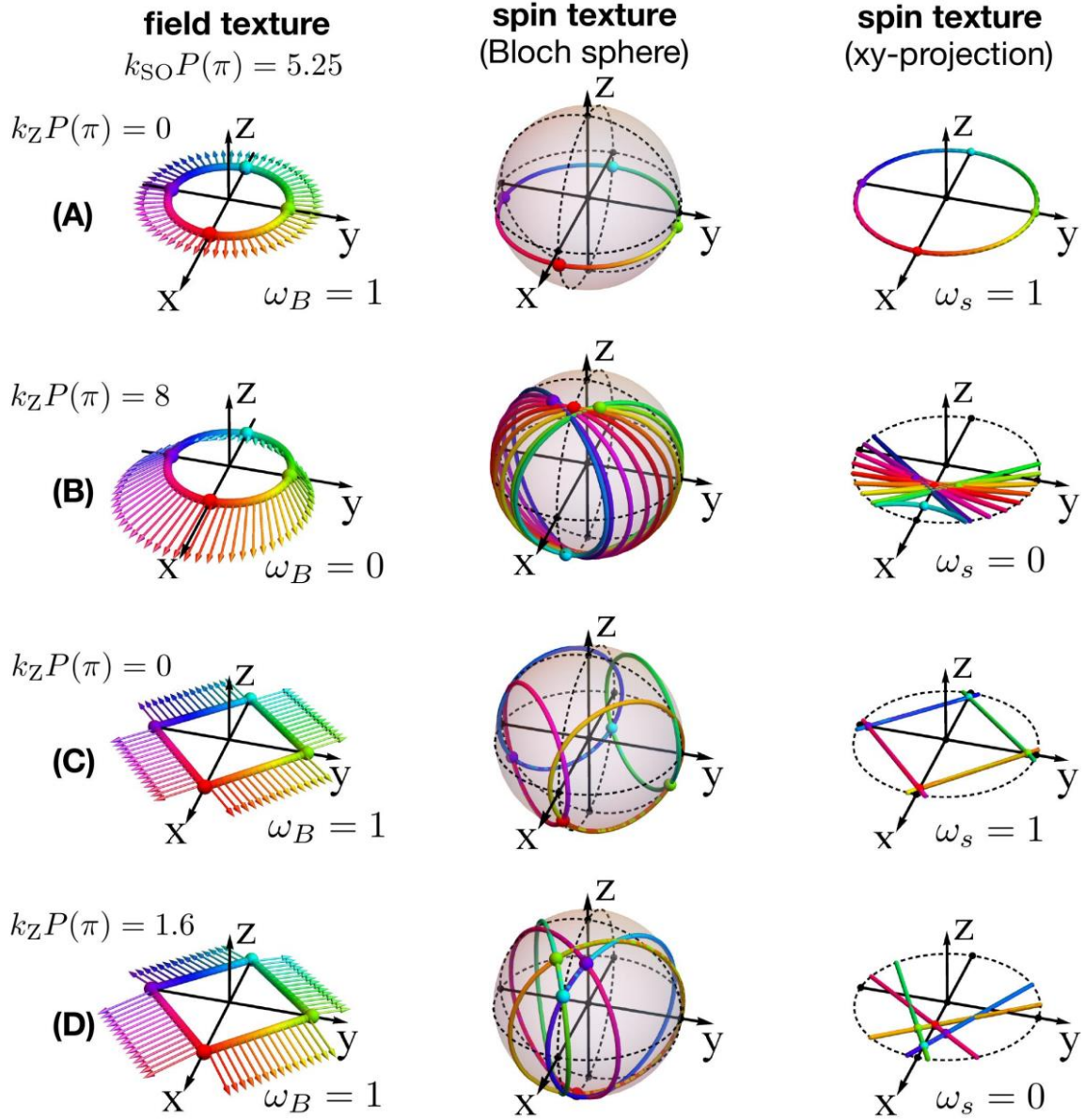

Fig. 2. Field and spin textures for hybrid Rashba/Zeeman ring and square loops corresponding to points (A), (B), (C) and (D) shown in Fig. 1. Field and spin topologies are defined by winding numbers $\omega_B$ and $\omega_s$ around the origin in the *xy*-plane. For rings, (A) and (B), a change in spin texture topology ($\omega_s$) requires large in-plane Zeeman fields changing the field topology ($\omega_B$) as well. For squares, (C) and (D), weak Zeeman fields are sufficient to change $\omega_s$ without changing $\omega_B$, instead. The Rashba SO components of the field textures correspond to counterclockwise itinerant spins, inverting its sign for clockwise spin carriers.

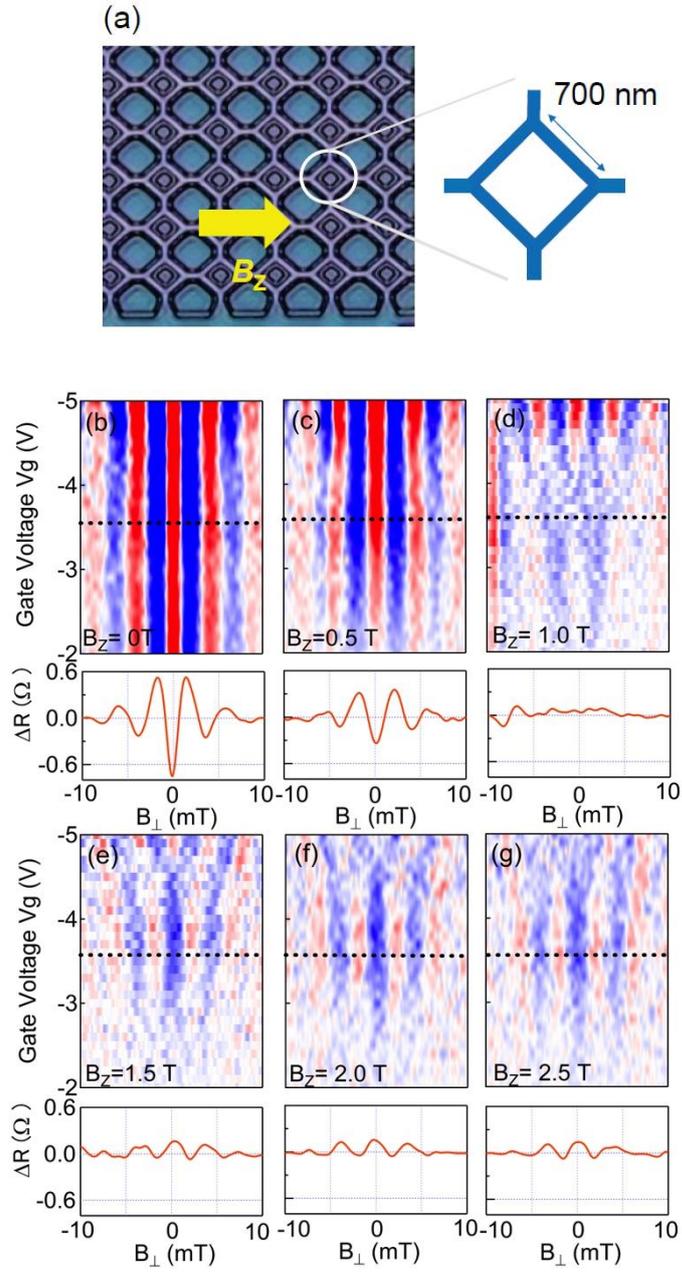

Fig. 3. (a) Laser scanning microscope image of a 40×40 square loop array with square side length $l = 700$ nm. (b)-(g) Contour color plots of MR as a function of $V_g$ at different in-plane field strengths $B_Z$. The insets of (b)-(g) contain the MR data at $V_g=$ -3.6 V, showing the AAS oscillations.

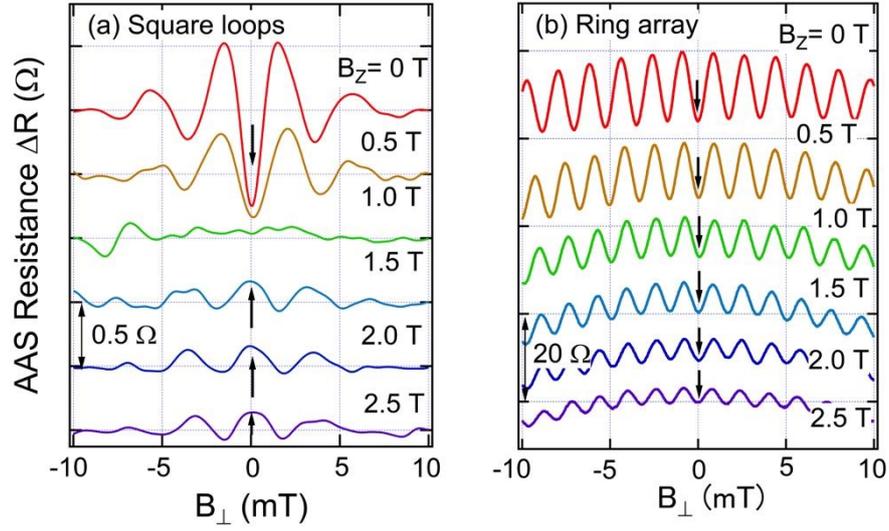

Fig. 4. (a) AAS oscillations for a square loop array at $V_g = -3.6$ V ( $\alpha = -2.5$ peVm ) with different in-plane Zeeman fields $B_Z$. The sign reversal of the AAS oscillations corresponds to the transition from a destructive spin interference (WAL) to a constructive one (WL). (b) AAS oscillations for a ring array with radius $r = 600$ nm at $\alpha = -2.9$ peVm. The transition from WAL to WL is not observed in the ring array.

## Supplementary Material

## Geometry-assisted topological transitions in spin interferometry

M. Wang, H. Saarikoski, A. A. Reynoso, J. P. Baltanás, D. Frustaglia, and J. Nitta

**Semiclassical theory of conductance:**

Here we elaborate on the one-dimensional (1D) semiclassical model for the two-contact quantum conductance of Rashba spin-orbit (SO) square- and ring-shaped loops used in our work. We adopt the Landauer-Büttiker formulation at zero temperature by identifying the linear conductance $G$ with the quantum transmission $T$ (in units of the quantum of conductance $e^2/h$):

$$G = \frac{e^2}{h} T, \qquad (S1)$$

where $T = \sum_{mn} |t_{mn}|^2$ and $t_{mn}$ is the quantum transmission amplitude from the incoming mode $n$ at the source contact lead to the outgoing mode $m$ at the drain contact lead. In our 1D model for spin-carrier transport we have only one orbital mode and two spin modes, such that $0 \leq T \leq 2$. Moreover, unitarity imposes $T + R = 2$, where $R = \sum_{mn} |r_{mn}|^2$ is the quantum reflection with $r_{mn}$ the corresponding amplitudes for incoming and outgoing modes $n$ and $m$ at the source contact lead. Moreover, in agreement with the experimental conditions, we work within the semiclassical regime [S1] where the carriers wavelength is much smaller than the system size and assume that the spin energy is much smaller than the kinetic energy (so that the spin dynamics does not alter the orbital one) [S2]. In this way, by following a path-integral approach and taking the semiclassical limit [S3], the quantum transmission amplitudes can be written as [S4]

$$t_{mn} = \sum_\Gamma a_\Gamma \, e^{ik_F L_\Gamma} \langle m|U_\Gamma|n\rangle, \qquad (S2)$$

namely, as a sum of contributions over classical paths $\Gamma$ of length $L_\Gamma$ that take the spin carriers from the entrance to the exit leads with statistical weight $a_\Gamma$. Within this picture, charge and spin contributions are clearly differentiated. The charge (propagating with Fermi wave number $k_F$) contributes with the action phase $e^{ik_F L_\Gamma}$. As for the spin, carriers entering the system with spin $n$ can leave it with spin $m$ according to the path-dependent spin evolution operator $U_\Gamma$, which is determined by the particular fields experienced by the spin carriers along the classical path $\Gamma$. The corresponding contribution is then given by the matrix element $<m|U_\Gamma|n>$.

As for the quantum transmission, it consists of probability terms of the form

$$|t_{mn}|^2 = \sum_{\Gamma,\Gamma'} a_\Gamma a_{\Gamma'}^* \, e^{ik_F(L_\Gamma - L_{\Gamma'})} \langle m|U_\Gamma|n\rangle \langle n|U_{\Gamma'}|m\rangle^*. \qquad (S3)$$

For a realistic modeling of the experimental conditions, the effects of disorder and/or sample averaging need to be taken into account. This means that (S3) needs to be averaged over different configurations including classical path fluctuations. Moreover, an average over a small energy window around the Fermi energy $E_F$ can be also implemented to take into account the effects of a finite (but low) temperature. Due to the presence of the orbital-phase factors $e^{ik_F(L_\Gamma - L_{\Gamma'})}$, the averaging procedure shows that the only surviving terms in (S3) are those corresponding to pairs of paths $\{\Gamma, \Gamma'\}$ with the same geometric length, $L_\Gamma = L_{\Gamma'}$. Other contributions simply average out due to rapid oscillations of the orbital-phase factors. However, identifying these pairs of paths contributing to the transmission is generally difficult. In this situation, it results most convenient to resort to the quantum reflection $R$ by taking advantage of unitarity. In analogy to (S3), the corresponding reflection probabilities read

$$|r_{mn}|^2 = \sum_{\Gamma,\Gamma'} b_\Gamma b_{\Gamma'}^* \, e^{ik_F(L_\Gamma - L_{\Gamma'})} \langle m|U_\Gamma|n\rangle \langle n|U_{\Gamma'}|m\rangle^*, \qquad (S4)$$

where the paths $\Gamma$ start and end at the source contact lead with statistical weight $b_\Gamma$. The averaging procedure in (S4) demands a pairing of paths with the same geometric length as well as in (S3). In contrast, finding these paths in (S4) is much easier once we notice that for any backscattering path $\Gamma$ there exists another path $\tilde{\Gamma}$ with exactly the same length that follows the trajectory defined by $\Gamma$ but starting and ending at opposite ends. Namely, $\Gamma$ and $\tilde{\Gamma}$ are time-reversal (TR) orbital paths (see Fig. S1).

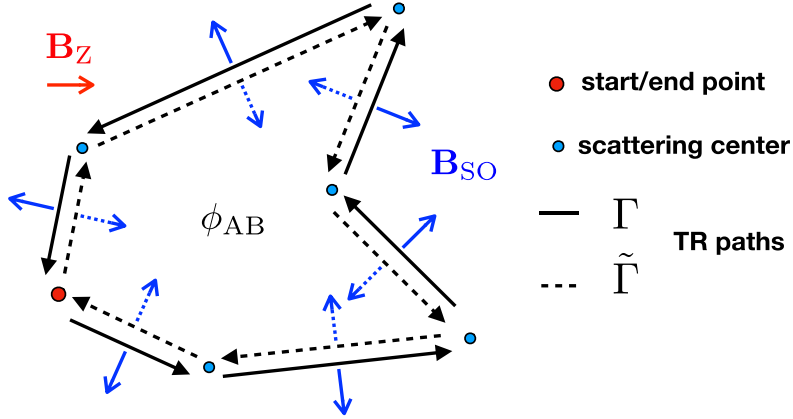

Fig. S1. Time-reversal (TR) paths dominating the contributions to the semiclassical conductance in two-dimensional (2D) disordered samples. In the presence of Rashba spin-orbit (SO) coupling, an effective (momentum-dependent) in-plane magnetic field $\boldsymbol{B}_{SO}$ shows up displaying complex field textures. These textures can be modified by introducing an additional in-plane Zeeman field $\boldsymbol{B}_Z$, with significant consequences on spin dynamics—see Eq. (S8). Moreover, by applying a perpendicular, weak magnetic field $B_\perp$, the paths enclose an Aharonov-Bohm (AB) magnetic flux $\phi_{AB}$ leading to Al'tshuler-Aronov-Spivak (AAS) oscillations in the conductance of mesoscopic loops.

As a consequence, the quantum conductance from TR contributing paths reads

$$G_{\text{TR-paths}} = \frac{e^2}{h}\left(2 - \text{tr}[\mathbb{R}\mathbb{R}^\dagger]\right) \qquad (S5)$$

where $\mathbb{R} = [r_{mn}]$ is a $2 \times 2$ spin matrix with elements

$$r_{mn} = \sum_\Gamma (b_\Gamma \langle m|U_\Gamma|n\rangle + b_{\tilde{\Gamma}} \langle m|U_{\tilde{\Gamma}}|n\rangle). \qquad (S6)$$

Hence, we find that the TR-path conductance given by (S5) and (S6) is fundamentally determined by the spin dynamics. The difficulty lies in finding the spin evolution operators $U_\Gamma$ along all possible paths $\Gamma$, which depends on the fields experienced by the carriers along the particular trajectories. Notice that in the absence of Rashba SO and Zeeman fields the spin evolution would be trivial ($U_{\Gamma/\tilde{\Gamma}} = \mathbb{I}_{2\times 2}$) and the conductance (S5) minimizes due to constructive TR-path interference. This effect is usually referred to as weak localization (WL). The introduction of a Rashba SO field can revert this effect. If the SO spin phases gathered along the TR paths are sufficiently large, the conductance (S5) can be maximized due to destructive TR-path interference leading to the so-called weak antilocalization (WAL) effect. We further notice that the introduction of a uniform Zeeman fields alone does not lead to WAL due to the "freezing" of spin dynamics (spin states quantize along the uniform field's axis all over the system).

**1D Rashba/Zeeman polygonal loops:**

We consider 1D ring- and polygon-shaped loop conductors symmetrically coupled to source and drain contact leads. In such doubly-connected geometries the classical paths are classified into clockwise (CW) and counterclockwise (CCW), with multiple possible windings around the loop. When the loops are well coupled to the leads, the carriers tend to escape the loop after a few windings, only. In this situation, all significant properties of the TR-path conductance are captured by considering exclusively single-winding paths [S5, S6]. This lowest-order model reduces the reflection amplitudes to

$$r_{mn} = \frac{1}{2}(\langle m|U_+|n\rangle + \langle m|U_-|n\rangle) = \frac{1}{2}\langle m|U_+ + U_-|n\rangle, \quad (S7)$$

where +/− label TR single-winding CCW/CW paths with geometrical length equal to the loop's perimeter $P$. We now proceed to calculate the spin evolution operators $U_{+/-}$ for polygon and ring loops subject to Rashba SO and Zeeman couplings.

We consider regular polygons lying on the $xy$-plane consisting of $N$ conducting segments of length $l = P/N$ connecting vertices $p$ and $q$ and oriented along directions $\hat{\gamma}$. The spin-carrier dynamics along each wire segment is determined by the Hamiltonian

$$H_\gamma = \frac{p_\gamma^2}{2m^*} + \frac{\alpha}{\hbar} p_\gamma (\hat{\gamma} \times \hat{z}) \cdot \boldsymbol{\sigma} + \frac{g\mu_B}{2} B_Z \sigma_x, \quad (S8)$$

with corresponding kinetic, Rashba SO, and Zeeman terms. Notice that the second terms in (S8) appears as an effective in-plane magnetic field $\boldsymbol{B}_{SO} = \frac{2\alpha}{\hbar g \mu_B} p_\gamma (\hat{\gamma} \times \hat{z})$, which is proportional to the momentum (inverting its sign for counter-propagating carriers) and perpendicular to the wire's direction $\hat{\gamma}$– see Figs. S1 and S2. In experiments, the Rashba coupling strength $\alpha$ is controlled electrically by means of top-gate voltages.

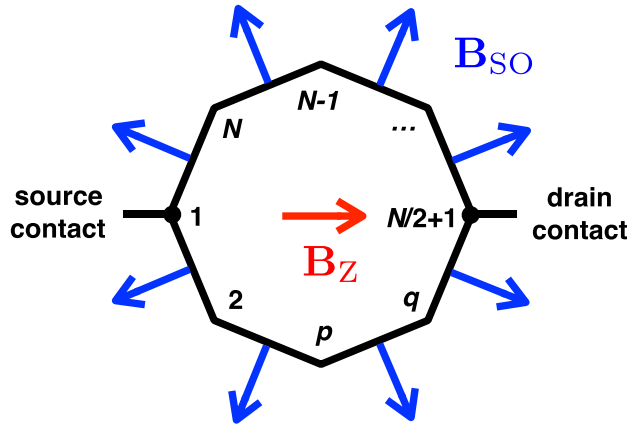

Fig. S2. 1D semiclassical model for the two-contact conductance of Rashba SO square- and ring-shaped loops used in our work. The Rashba field $\boldsymbol{B}_{SO}$ correspond to CCW propagating spin carriers.

The spin dynamics of a carrier propagating along the segment connecting the vertices $p$ and $q$ is described by a spin-rotation operator

$$R_{q,p} = \exp\left[-i\frac{t_N}{\hbar}\frac{g\mu_B}{2}\boldsymbol{B}\cdot\boldsymbol{\sigma}\right], \quad (S9)$$

which accounts for the spin precessions around the effective in-plane magnetic field, $\boldsymbol{B} = \boldsymbol{B}_{SO} + \boldsymbol{B}_Z$, due to the combined action of the local ($\hat{\boldsymbol{\gamma}}$- and $qp$-dependent) Rashba SO field and the uniform Zeeman field during a fight time $t_N = l/v_F$, with $v_F$ the Fermi velocity. In this way, the spin evolution operators along TR paths in (S7) reduce to [S7]

$$U_+(N) = R_{1,N}...R_{3,2}\, R_{2,1},$$
$$U_-(N) = R_{1,2}...R_{N-1,N}\, R_{N,1}, \quad (S10)$$

and the reflection matrix used to calculate the TR-path conductance in (S5) reads

$$\mathbb{R}(N) = \tfrac{1}{2}[U_+(N) + U_-(N)]. \quad (S11)$$

Notice that the carriers spin textures around a loop can be obtained by (i) calculating the eigenvectors of $U_{+/-}(N)$ and (ii) propagating them along the loop's perimeter accordingly. This is how we obtained the spin textures of Fig. 2.

**Results for 1D square and ring loops:**

In our 1D simulations, the Rashba SO and Zeeman coupling strengths are naturally quantified in terms of the corresponding spin precession angles during propagation along the loop's perimeter, $k_{SO}P$ and $k_ZP$, with $k_{SO} = \frac{\alpha m^*}{\hbar^2}$, $k_Z = \frac{g\mu_B B_Z m^*}{2\hbar^2 k_F}$, and $k_F$ the Fermi wave number.

We start by discussing the case of ring geometries. Rings can be studied by taking the limit of large $N$ in Eqs. (S9)-(S11). In practice, this means that the spin precession length must be much larger that the polygon sides $l$, implying $N \gg k_{SO}P/\pi, k_ZP/\pi$. The simulations of Figs. 1(a), 1(c), 2(A) and 2(B) were performed with $N = 512$.

An analytical expression for the TR-path conductance of rings can be obtained in the limit of weak Zeeman fields ($k_ZP \ll k_{SO}P$) and near adiabatic spin dynamics ($k_{SO}P/\pi \gg 1$, where the spin textures are close to the equator of the Bloch sphere). There we find [S6]

$$G_{\text{ring}} = \frac{e^2}{h}\left[1 - \cos(2k_{SO}P + \phi_g)\right],$$
$$\phi_g = \frac{(k_ZP)^2}{2k_{SO}P}. \quad (S12)$$

The ring's conductance (S12) displays Aharonov-Casher (AC) oscillations as a function of $k_{SO}P$ with period 1 in $\pi$ units. These oscillations are illustrated along the horizontal axis of Fig. 1(a). In this regime, the spin textures describe cone-like trajectories in the Bloch sphere winding around the $z$-axis, see Figs. 1(c) and 2(A). The introduction of the in-plane Zeeman field produces a phase shift $\phi_g$ of geometric origin (as demonstrated in [S6]), leading to the bent, vertical interference bands shown in Fig. 1(a). The Zeeman field distorts the spin textures (modifying the subtended solid angles in the Bloch sphere and, consequently, the associated geometric phases) without modifying its topological characteristics in terms of windings around the $z$-axis. Large Zeeman fields ($k_Z P > k_{SO} P$) modifying the topology of the effective-field texture $\boldsymbol{B}$ are necessary to change the spin texture's winding, as seen in Figs. 1(c) and 2(B).

As for square loops, an exact solution for the TR-path conductance can by obtained in the absence of Zeeman fields:

$$G_{\text{square}}(B_Z = 0) = \frac{e^2}{h}\left\{2 - \frac{1}{2}\left[2 - \left(1 - \cos\left(\frac{k_{SO}P}{2}\right)\right)^2\right]^2\right\}. \qquad (S13)$$

A close look at the square's conductance (S13) shows that it presents AC oscillations as a function of $k_{SO}P$ with period 2 in $\pi$ units. This period doubling with respect of the ring's conductance (S12)— observed along the horizontal axis in Fig. 1(b)— is a consequence of a strongly non-adiabatic spin dynamics due to the Rashba field discontinuities at the square's corners. This keeps the spin textures away from the field's plane as shown in Fig. 2(C), hindering the gathering of AC phases (proportional to the local spin/field projection). As the Zeeman field is introduced, an approximate analytical expression for the TR-path conductance can be found in the limit of large Rashba strengths, $k_{SO}P/\pi \gg 1$, which in the case of squares does not correspond to an adiabatic regime by effect of the corners. Within this approximation, illustrated in Fig. S3, we find

$$G_{\text{square}} = \frac{e^2}{h}\left\{2 - 4\left[\cos\left(\frac{k_{SO}P}{4} - \frac{k_Z P}{4\sqrt{2}}\right)\cos\left(\frac{k_{SO}P}{4} + \frac{k_Z P}{4\sqrt{2}}\right)\sin\left(\frac{k_Z P}{2\sqrt{2}}\right)\right]^2 - $$

$$\frac{1}{2}\left[\cos 2\left(\frac{k_{SO}P}{4} - \frac{k_Z P}{4\sqrt{2}}\right) + \cos 2\left(\frac{k_{SO}P}{4} + \frac{k_Z P}{4\sqrt{2}}\right) + \sin 2\left(\frac{k_{SO}P}{4} - \frac{k_Z P}{4\sqrt{2}}\right)\sin 2\left(\frac{k_{SO}P}{4} + \frac{k_Z P}{4\sqrt{2}}\right)\right]^2 - $$

$$2\left[\sin^2\left(\frac{k_{SO}P}{4} + \frac{k_Z P}{4\sqrt{2}}\right) - \sin^2\left(\frac{k_{SO}P}{4} - \frac{k_Z P}{4\sqrt{2}}\right)\right]^2\right\}. \qquad (S14)$$

The (anti)symmetric combinations of $k_{SO}P$ and $k_Z P/\sqrt{2}$ in (S14) determine a checkerboard-like pattern for the conductance as depicted in Fig. 1(b), where a geometrical factor $1/\sqrt{2}$ arises due to the diagonal application of the Zeeman field. Figure S3 already indicates that the Zeeman field acts on the spin carriers on a strength scale similar to that of the Rashba field (since it manifests as a simple shift on the local Rashba fields), as confirmed by Eq. (S14).

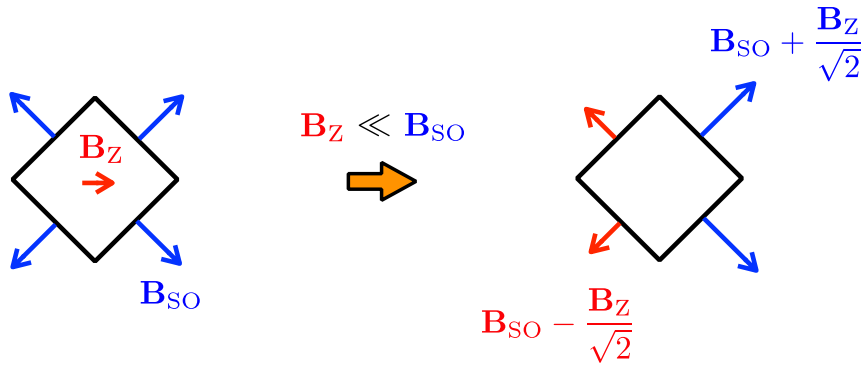

Fig. S3. Hybrid Rashba/Zeeman field texture for 1D squares in the limit of small Zeeman fields. The figure on the right suggests that, in that limit, Rashba and Zeeman fields act on the spin carriers on similar strength scales, as confirmed by the checkerboard-like pattern of Fig. 1(b). The Rashba field $\boldsymbol{B}_{SO}$ corresponds to CCW propagating spin carriers.

For a vanishing Zeeman field, as the Rashba strength increases the spin textures in squares can develop complex patterns on the Bloch sphere without changing its topological characteristics in terms of windings around the *z*-axis: they show a constant, finite winding number (coincident with that of the field texture) all over the Rashba axis as shown in Figs. 1(d) and 2(C). However, the introduction of a weak Zeeman field produces a topological transition in the spin textures by modifying the winding number as shown in Figs. 1(d) and 2(D). In contrast to rings, this happens without modifying the topology of the field texture: the strongly non-adiabatic spin dynamics in squares makes it easy for weak Zeeman field to produce a significant distortion on the spin textures without much effort, introducing a decorrelation between spin and field textures that prevents the gathering of spin phases contributing to WAL. This is why WL/WAL transitions driven by the Zeeman field act as topological indicators for spin textures in square loops, as discussed in Figs. 1(b), 1(d), 2(C) and 2(D).

**Al'tshuler-Aronov-Spivak (AAS) oscillations:**

By introducing a weak magnetic field $B_\perp$ perpendicular to the loop's plane, the TR paths contributing to the conductance enclose a magnetic Aharonov-Bohm (AB) flux $\phi_{AB} = B_\perp S$, where $S$ in the loop's area. The flux manifests as an additional contribution to the action phases in Eq. (S4) of the form $e^{ik_F L_\Gamma \pm 2\pi(\phi_{AB}/\phi_0)}$, where the sign depends on the CW/CCW circulation of the path $\Gamma$ and $\phi_0$ is the flux quantum. This modifies the reflection amplitudes (S7) by introducing corresponding phase prefactors:

$$r_{mn} = \frac{1}{2} \langle m | e^{i2\pi(\phi_{AB}/\phi_0)} U_+ + e^{-i2\pi(\phi_{AB}/\phi_0)} U_- | n \rangle, \quad (S15)$$

that eventually modulate the TR-path conductances of rings (S12) and squares (S14) as follows:

$$G_{\text{ring}} = \frac{e^2}{h} \left[ 1 - \cos\left(4\pi \frac{\phi_{AB}}{\phi_0}\right) \cos(2k_{SO} P + \phi_g) \right], \quad (S16)$$

$$G_{\text{square}} = \frac{e^2}{h} \left\{ 2 - \cos\left(4\pi \frac{\phi_{AB}}{\phi_0}\right) \left\{ 4 \left[ \cos\left(\frac{k_{SO} P}{4} - \frac{k_Z P}{4\sqrt{2}}\right) \cos\left(\frac{k_{SO} P}{4} + \frac{k_Z P}{4\sqrt{2}}\right) \sin\left(\frac{k_Z P}{2\sqrt{2}}\right) \right]^2 + \right.$$

$$\frac{1}{2} \left[ \cos 2\left(\frac{k_{SO} P}{4} - \frac{k_Z P}{4\sqrt{2}}\right) + \cos 2\left(\frac{k_{SO} P}{4} + \frac{k_Z P}{4\sqrt{2}}\right) + \sin 2\left(\frac{k_{SO} P}{4} - \frac{k_Z P}{4\sqrt{2}}\right) \sin 2\left(\frac{k_{SO} P}{4} + \frac{k_Z P}{4\sqrt{2}}\right) \right]^2 +$$

$$\left. 2 \left[ \sin^2\left(\frac{k_{SO} P}{4} + \frac{k_Z P}{4\sqrt{2}}\right) - \sin^2\left(\frac{k_{SO} P}{4} - \frac{k_Z P}{4\sqrt{2}}\right) \right]^2 \right\} \right\}. \quad (S17)$$

The TR-path conductances (S16) and (S17) oscillate as a function of the flux $\phi_{AB}$ with a period $\phi_0/2$, i.e., by showing a period halving with respect to the usual AB effect. These particular modulations of the conductance of mesoscopic loops by a magnetic flux due to TR-path interference are known as the AAS oscillations [S8].

In experiments, the existence of AAS oscillations demonstrates that TR paths contribute significantly to transport. Moreover, notice in (S16) and (S17) that the sign of the AAS oscillations is exclusively determined by the spin dynamics: conductance AAS oscillations starting with a minimum/maximum at $\phi_{AB} = 0$ is indicative of WL/WAL. This relation is reversed for the resistance (most commonly used when reporting experimental results): resistance AAS oscillations starting with a minimum/maximum at $\phi_{AB} = 0$ is indicative of WAL/WL.

**AC spin interference in experiments with square loops:**

Figure S4 shows contour color plots of the magnetoresistace (MR) as a function of the gate voltage $V_g$ without in-plane Zeeman field. The bottom inset of Fig. S4 contains the MR data at $V_g = $ -3.6 V, showing the AAS oscillations [S8] with a period $\Delta B_\perp = (h/2e)l^2$. The MR amplitude at $B_\perp = 0$ as a function of $V_g$ corresponds to the TR-AC spin interference. The AC modulations are seen in the right inset of Fig. S4. A full period is not covered because the AC oscillation period in the square loop is two times larger than the AC period of the ring-shaped spin interferometer ( -1.17 peVm $< \alpha <$ -3.62 peVm). In Fig. S5, we plot the TR-AC spin interference as a function of in-plane Zeeman field $B_Z$. The dip in the AC modulation turns into a peak by increasing $B_Z$, showing the transition from WAL to WL.

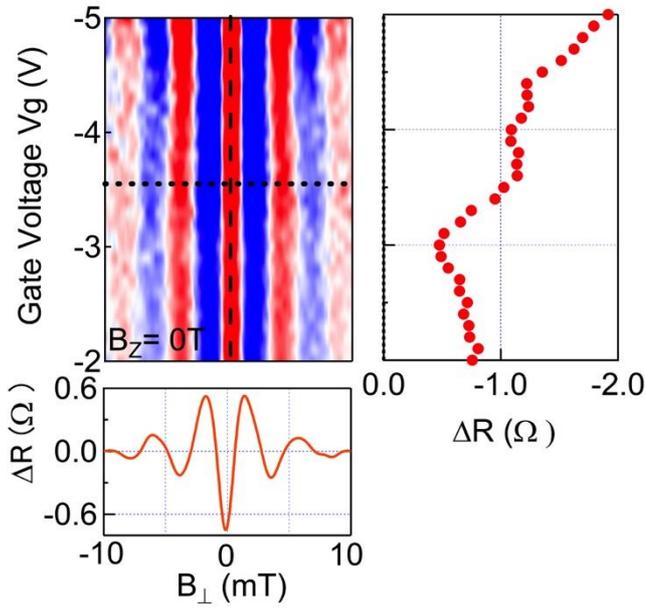

Fig. S4 Contour color plots of magnetoresistace (MR) as a function of the gate voltage $V_g$ without in-plane Zeeman field. The bottom inset contains the MR data at $V_g = $ -3.6 V, showing the AAS oscillations with a period $\Delta B_\perp = (h/2e)l^2$. The right inset shows the MR amplitude at $B_\perp = 0$ as a function of $V_g$, corresponding to the TR-AC spin interference.

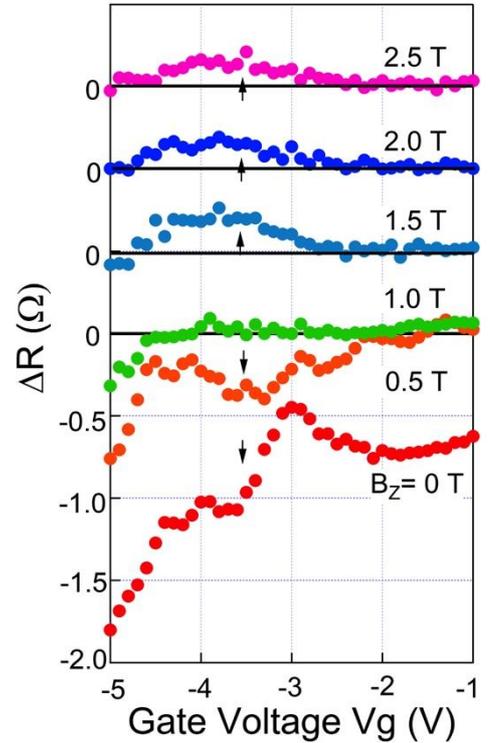

Fig. S5 TR-AC spin interference at $B_\perp = 0$ as a function of in-plane Zeeman field $B_Z$.

**Two-dimensional (2D) numerical simulations of transport:**

We perform 2D numerical simulations of disordered multi-mode square loops accounting for more realistic conditions, complementary to our 1D semiclassical model. Our 2D simulations rely on a tight-binding model solved by using a recursive Green's function method and the Kwant code [S9] with lattice spacing ~ 1 nm, five propagating modes, and a mean-free path of 1.2 μm (which is shorter that the square perimeter $P = 2.8$ μm). This disorder is crucial to develop TR-paths for AAS interference [S8]. The calculation details are described in [S10]. We assume a Rashba SO strength of $\alpha$= -2.0 peVm, and $g = 4$. Each side of the square loop is set to be 700 nm. These parameters are very similar to those in the InGaAs quantum well used in our experiment. The numerical result shown in Fig. S6(a) indicates that the reversal of the AAS interference appears around $B_Z$= 1.25 T, reproducing the experimental results of Fig. 4(a) rather well.

We notice that the critical $B_Z$ producing the topological transition depends on the mean free path as shown in Fig. S6(b). This finding is compatible with previous results showing that disorder contributes to the development of non-adiabatic spin dynamics [S11], boosting the effects of SO-field discontinuities at the square's corners.

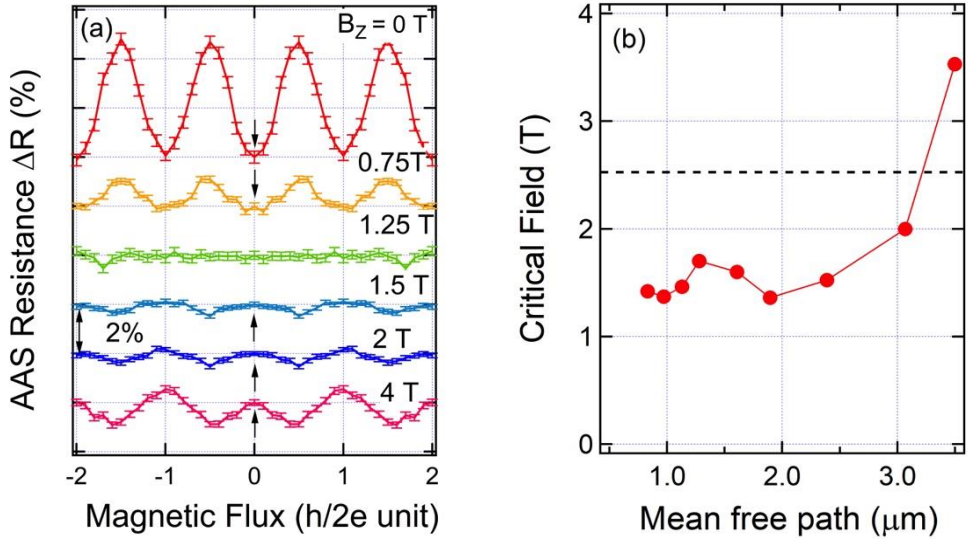

Fig. S6. (a) AAS oscillations for a square loop with side length $l = 700$ nm calculated by using 2D numerical simulations (Kwant code) with a mean-free path of 1.2 μm, Rashba SO strength $\alpha$= -2.0 peVm, and g-factor g= 4. (b) The critical in-plane field producing the AAS sign reversal (identified with a topological spin transition) depends on the mean free path. Other parameters are the same as in previous calculations. The dashed line corresponds to the critical field found with our 1D TR-path model, Fig. 1(b).